\documentclass[11pt]{article}
\usepackage{latexsym}
\usepackage{amssymb}
\usepackage{epsf}
\usepackage{amsmath}
\usepackage{slashed}
\usepackage[hypertex]{hyperref}
\usepackage{graphicx}

\newcommand{\tr}{{\rm Tr}}

\begin{document}
\pagestyle{empty}

\begin{flushright}
TU-876
\end{flushright}

\vspace{3cm}

\begin{center}

{\bf\LARGE SuperTopcolor}
\\

\vspace*{1.5cm}
{\large 
Hiraku Fukushima, Ryuichiro Kitano and Masahiro Yamaguchi
} \\
\vspace*{0.5cm}

{\it Department of Physics, Tohoku University, Sendai 980-8578, Japan}\\
\vspace*{0.5cm}

\end{center}

\vspace*{1.0cm}

\begin{abstract}
{\normalsize
We consider a supersymmetric QCD with soft supersymmetry breaking terms
 as the dynamics for the electroweak symmetry breaking.
 We find various advantages compared to the non-supersymmetric models,
 such as a natural incorporation of the dynamical top-quark mass
 generation (the topcolor mechanism), the existence of a boson-pair
 condensation (the composite Higgs fields) and a large anomalous
 dimension of the composite operator to cure the
 flavor-changing-neutral-current and the $S$-parameter crises of the
 technicolor theories. 
The knowledge of the weakly coupled description (the Seiberg duality)
 enables us to perform perturbative computations in strongly coupled
 theories.
Working in a large flavor theory where perturbative calculations are
 reliable in the dual description, one can find a stable vacuum with chiral
 symmetry breaking. The top/bottom quarks and also the Higgsinos obtain
 masses through a dynamically generated superpotential.

}
\end{abstract} 

\newpage
\baselineskip=18pt
\setcounter{page}{2}
\pagestyle{plain}
\baselineskip=18pt
\pagestyle{plain}

\setcounter{footnote}{0}

\section{Introduction}

Technicolor is a conceptually attractive scenario for electroweak
symmetry breaking. A strong dynamics deforms the vacuum at an infrared
energy scale, and the particles living on the vacuum obtain masses
through the deformation~\cite{Weinberg:1975gm,Susskind:1978ms}. Indeed,
the QCD is operating as a minor source of the masses of the $W$ and $Z$
bosons. It sounds natural to consider that there is main dynamics for
the electroweak symmetry breaking instead of anticipating an elementary
Higgs field.

Usually in QCD-like technicolor theories, the Standard Model fermions
obtain masses not directly through the dynamics; they are treated as
massless until the extended technicolor (ETC)
interactions~\cite{Dimopoulos:1979es, Eichten:1979ah} are turned on. It
is then a question that why the top quark is so heavy. This question
suggests to consider a model in which the top quark is more involved in
the technicolor dynamics.
The topcolor model~\cite{Hill:1991at} takes an interesting (the most
radical) approach that the top quark plays the leading role in the
electroweak symmetry breaking; the symmetry breaking by the condensation
of the top-quark pair, $\langle \bar t t \rangle \neq
0$~\cite{Miransky:1988xi,Bardeen:1989ds}.

The essence of the topcolor model is the following. The top quark
carries a color of the $SU(3)_{TC}$ gauge interaction, instead of that
of QCD. By the strong $SU(3)_{TC}$ gauge dynamics, a condensation
$\langle \bar t t \rangle$ is formed and breaks electroweak gauge
symmetry just as in the chiral symmetry breaking in QCD.
The big difference from QCD is the assumption that the top-quark degree
of freedom somehow remains as a massive asymptotic state while the
topcolor quantum number is replaced by the color in QCD.
It is assumed that this can be realized by a gauge symmetry breaking,
$SU(3)_{TC}$ $\times$ $SU(3)_{C^\prime}$ $\to$ $SU(3)_C$, where the
unbroken gauge group is the QCD.

Although it is an interesting plot, we need more ingredients for this
model to be a complete framework. For example, we need a mechanism for
other fermion masses. A simple extension with an ETC sector would result
in too large flavor changing neutral current (FCNC) processes.
The electroweak precision tests also put severe constraints on QCD-like
theories.
The gauge symmetry breaking, $SU(3)_{TC}$ $\times$ $SU(3)_{C^\prime}$
$\to$ $SU(3)_C$, requires yet another dynamics or a Higgs field in the
full theory.
Moreover, a naive estimation of the top quark mass seems to give a too
large value compared to the observed one. As an approach to a realistic
model, for example, it has been proposed to combine the technicolor
(with ETC) and the topcolor: the topcolor assisted technicolor
model~\cite{Hill:1994hp}. It has also been proposed to introduce a Higgs
field in the topcolor model~\cite{Aranda:2000vk}.
See \cite{Hill:2002ap} for a review of the technicolor and topcolor
models.

In this paper, we take a different approach to this series of problems
in the dynamical electroweak symmetry breaking.
We consider a softly broken supersymmetric (SUSY) QCD as the dynamical
model for the electroweak symmetry breaking.
The topcolor model is naturally incorporated in the framework. In
particular, the topcolor dynamics itself works as a mechanism to convert
the topcolor of the top quark into the color in QCD.
We here list advantages compared to non-SUSY topcolor models;
\begin{itemize}
 \item The minimal model turns out to have an infrared fixed points,
       {\it i.e.,} within the conformal
       window~\cite{Seiberg:1994pq}. The existence of an energy regime
       of a scale-invariant theory has a virtue of alleviating the FCNC
       problem~\cite{Holdom:1981rm, Yamawaki:1985zg, Akiba:1985rr,
       Appelquist:1986an, Luty:2004ye}.  In addition in SUSY models, the
       existence of a boson-pair condensation further helps the
       situation~\cite{Buras:1982nb}. The dimension of the composite
       Higgs operator (the stop-pair condensation) is $D(H) = 1.2$ with
       which the ETC scale can be postponed up to $10^{12}$~GeV.
 \item The model is at the most strongly coupled edge of the conformal
window, at which the theory would be out of control if it was not
       supersymmetric. However, thanks to the Seiberg
       duality~\cite{Seiberg:1994pq}, there is a weakly coupled
       description in terms of a meson and dual-quark superfields. As
       components of the meson superfields, one can find the top and
       bottom quarks where the the topcolor quantum number is already
       replaced by the ordinary color.
 \item The Higgs superfields are present in the dual theory as
       components of the meson superfields. The electroweak symmetry
       breaking is described by the Higgs mechanism, where the Higgs
       potential is generated by non-perturbative effects of the dual
       gauge dynamics and also from the soft SUSY breaking terms. In a
       large $N_f$ theory where the perturbative expansion is reliable,
       one can find a stable vacuum with non-vanishing vacuum
       expectation values (VEVs).
 \item The non-perturbatively generated superpotential produces the
       masses of the top/bottom quarks and the Higgsino fields.
 \item A perturbative calculation of the $S$ parameter is possible in
       the dual theory. We find that the contribution from the
       dynamical sector (the dual quark loops) is small enough, $\Delta
       S \sim 0.1$.
\end{itemize}

The overall picture looks nice, but we learn that $N_f = 5$ in the
actual model is not quite large enough to make precise predictions on
the mass spectrum. 
With a small $N_f$, the gauge interaction is strong enough that the size
of the non-perturbative effects becomes comparable to the soft-SUSY
breaking terms. The exact superpotential cannot be reliably used in such
a situation.
We assume, however, that the qualitative picture of a large $N_f$ theory
remains true for $N_f = 5$. In this case, the low energy effective
theory looks like the minimal SUSY Standard Model (MSSM) but the
$\mu$-term and the top/bottom Yukawa interactions replaced by a
dynamical superpotential, where fractional powers of the superfields
appear as a consequence of integrating out the dual quarks.
Therefore, the phenomenology of the Higgs fields and the top/bottom
quarks will be different from that of the MSSM.

One can view the dual model directly as an extension of the MSSM. Unlike
the MSSM, the lightest Higgs boson mass is not related to the gauge
coupling constant; rather the main contribution is the non-perturbative
effect of the dual gauge interaction. Therefore, the Higgs boson can
naturally be heavy enough to evade the lower bound from the LEP-II
experiments~\cite{Barate:2003sz}. Also, as we will see later, the soft
masses for the Higgs fields do not receive a correction from the top
Yukawa interaction. This is in fact a desired situation in terms of the
fine-tuning problem in SUSY models. (See \cite{Kobayashi:2005mg,
Kobayashi:2006fh} for a related proposal to realize a small top Yukawa
coupling constant at high energy.)

The combination of SUSY with technicolor has been studied in the
literature.
There is a paper ``Topcolor-Assisted Supersymmetry''~\cite{Liu:1999ah}
which discussed a use of topcolor in gauge mediation. Although it sounds
similar, our approach is quite different.
Our approach is more related to the supersymmetric technicolor model
proposed in \cite{Luty:2000fj, Murayama:2003ag} and the fat Higgs/top
models in~\cite{Harnik:2003rs} and \cite{Delgado:2005fq}.
These models assume the dynamical electroweak symmetry breaking in the
supersymmetric limit.  In contrast, we consider a model where the
SUSY-breaking terms trigger the electroweak symmetry breaking.
Historically, the supersymmetric technicolor is proposed in
\cite{Witten:1981nf} and models are constructed in
Refs.~\cite{Dine:1981za,Dimopoulos:1981au}.

\section{Model}

We consider a SUSY QCD model with $N_c = 3$ and $N_f = 5$. The $SU(N_c)$
gauge interaction is the topcolor gauge group, $SU(3)_{TC}$, which gets
strongly coupled at a low energy scale.
The particle content and the quantum numbers are listed in
Table~\ref{tab:ele}.
The chiral superfields $Q$ and $\bar Q$ belong to the fundamental and
the anti-fundamental representations of $SU(N_c)$.
The Standard Model gauge group ($SU(3)_C$ $\times$ $SU(2)_L$ $\times $
$U(1)_Y$) is embedded in the flavor symmetry $SU(N_f)_L$ $\times$
$SU(N_f)_R$ $\times$ $U(1)_B$ such that the quantum numbers of $Q$ and
$\bar Q$ are given by
\begin{eqnarray}
\renewcommand{\arraystretch}{1.2}
 Q = \left(
\begin{array}{c}
 (1,2)_{1/6}\\ \hline
 (\bar 3, 1)_0\\
\end{array}
\right),\ \ 
 \bar Q = \left(
\begin{array}{c}
 (1,1)_{-2/3}\\
 (1,1)_{1/3}\\ \hline
  (3, 1)_0\\
\end{array}
\right).
\label{eq:qqbar}
\renewcommand{\arraystretch}{1}
\end{eqnarray}
The first and the second entries corresponds to the $SU(3)_C$ and
$SU(2)_L$ quantum numbers, respectively, and the subscripts are the
$U(1)_Y$ hypercharges.
The upper two flavors correspond to the top and bottom superfields where
the Standard Model color is replaced by the topcolor $SU(3)_{TC}$. The
lower three flavors are bi-fundamental fields charged under both the
topcolor and the Standard Model color group.
Together with other chiral superfields (all the lepton superfields and
the quark superfields in the first and the second generations), the
theory is free from gauge anomaly.
This cancellation of the gauge anomaly is one of the attractive features
of the topcolor model. Just replacing the color of top and
bottom quarks by topcolor does not cause gauge anomaly. 

\renewcommand{\arraystretch}{1.3}
\begin{table}[t]
\begin{center}
 \begin{tabular}[t]{c|ccccc}
  & $SU(N_c)$ & $SU(N_f)_L$ & $SU(N_f)_R$ & $U(1)_B$ & $U(1)_R$ \\ \hline
 $Q$ & $N_c$ & $N_f$ & $1$ & 1 & $(N_f - N_c)/N_f$ \\
 $\overline Q$ & $\overline {N_c}$& 1 & $\overline {N_f}$ & $-1$ & $(N_f - N_c)/N_f$ \\ 
 \end{tabular}
\end{center}
\caption{Quantum numbers of the SUSY QCD.}
\label{tab:ele}
\end{table}
\renewcommand{\arraystretch}{1}

In the original topcolor model in Ref.~\cite{Hill:1991at}, the fermionic
components of upper two flavors and bosonic components of lower three
flavors are introduced.
It is assumed that the bi-fundamental scalar field obtains a VEV to
break the $SU(3)_{C^\prime}$ $\times$ $SU(3)_{TC}$ gauge group down to the
diagonal $SU(3)$ group so that the top quark carries the ordinary color
at low energy.
Under the $SU(3)_{TC}$ dynamics (although broken), the top quark is
assumed to obtain a dynamical mass and also the top-quark condensation
is formed to break the $SU(2)_L$ $\times$ $U(1)_Y$ gauge group.
We will see these phenomena more explicitly in this SUSY version of the
model.

For $N_c = 3$ and $N_f = 5$, the model is in the conformal
window. Therefore, without any perturbation, the theory flows into the
non-trivial infrared (IR) fixed point at some scale and stays as a
conformal field theory until mass parameters in the theory get
important.

Since we anyway need SUSY breaking, we add soft SUSY breaking terms to
the Lagrangian, such as the gaugino masses and the scalar masses with
typical sizes $m_{\rm SUSY}$ of the order of the electroweak scale. We
assume that the theory first approaches the IR fixed point at a scale
$\Lambda$ $(\gg m_{\rm SUSY})$. At a low energy scale where the soft
terms become important, one can expect that the topcolor mechanism
works, {\it i.e.}, the dynamical electroweak symmetry breaking occurs.

For the analysis of this model, the Seiberg duality is very
powerful. Since $N_f = 5$ is the most strongly coupled edge of the
conformal window, the dual picture provides us with a weakly coupled
description in terms of dual quarks and a meson superfield.
By the assumption of $\Lambda \gg m_{\rm SUSY}$, the Seiberg duality can
be reliably used for the analysis~\cite{Aharony:1995zh}.
The dual theory is also a gauge theory with soft SUSY breaking terms.
We proceed to the discussion of the dual description in the next
section. For a while, we ignore the Standard Model gauge interactions.

\section{Dual model and a large $N_f$ expansion}

\renewcommand{\arraystretch}{1.3}
\begin{table}[t]
\begin{center}
 \begin{tabular}[t]{c|ccccc}
  & $SU(N)$ & $SU(N_f)_L$ & $SU(N_f)_R$ & $U(1)_B$ & $U(1)_R$ \\ \hline
 $M$ & $1$ & $N_f$ & $\overline {N_f}$ & 0 & $2N/N_f$ \\ $q$ & $N$&
 $\overline{N_f}$ & 1 & $N_c/N$ & $N_c/N_f$ \\ $\bar q$ & $\overline N$
 & $1$ & $N_f$ & $-N_c/N$ & $N_c/N_f$ \\
 \end{tabular}
\end{center}
\caption{Particle content and quantum numbers in the dual picture.}
\label{tab:dual}
\end{table}
\renewcommand{\arraystretch}{1}

The dual theory is a $SU(N)$ gauge theory with $N= N_f - N_c$. The
particle content and the quantum numbers are listed in
Table~\ref{tab:dual}.  In this picture, the top/bottom quarks and the
Higgs fields can be found in the $SU(N)$ singlet meson superfield $M$:
\begin{eqnarray}
 M = \left(
\begin{array}{cc|c}
 H_d & H_u & q_3 = 
\left(
\begin{array}{c}
 t \\
 b \\
\end{array}
\right)
 \\ \hline
 t^c & b^c& X  \\
\end{array}
\right).
\end{eqnarray}
The $X$ superfield is composed of a singlet, $X_{\bf 1}$, and
an adjoint field, $X_{\rm adj}$, under $SU(3)_C$.
Here the top and bottom quark superfields, $q_3$, $t^c$, and $b^c$, are
now charged under $SU(3)_C$, instead of the topcolor $SU(3)_{TC}$.

The dual quarks $q$ and $\bar q$ are fundamental and anti-fundamental
under the dual gauge group, respectively, and have the following
Standard Model quantum numbers:
\begin{eqnarray}
 \renewcommand{\arraystretch}{1.2}
 q = \left(
\begin{array}{c}
 (1,2)_{0}\\ \hline
 (3, 1)_{1/6}
\end{array}
\right),\ \ 
 \bar q = \left(
\begin{array}{c}
 (1,1)_{1/2}\\ 
 (1,1)_{-1/2}\\ \hline
 (\bar 3,1)_{-1/6}
\end{array}
\right).
\end{eqnarray}
Interestingly, this dual picture has a structure of the MSSM plus the
sector of the minimal (walking) technicolor~\cite{Weinberg:1975gm,
Susskind:1978ms} where the technicolor group is $SU(2)$. (See
Ref.~\cite{Sannino:2004qp} for the minimal walking technicolor
model. The model contains a fermion in the two-index symmetric
representation which can be identified as the dual gaugino in the
present model.)

The superpotential is
\begin{eqnarray}
 W = h q M \bar q,
\end{eqnarray}
with $h$ the coupling constant.
Once we generalize $N_f=5$ to a large value with $3N-N_f$ ($=2N_f-3N_c$)
fixed to be unity, the model is at the Banks-Zaks fixed
point~\cite{Banks:1981nn,Seiberg:1994pq} where one can do a perturbative
expansion. By using a relation between the $R$-charge and the dimension
of chiral operators,
\begin{eqnarray}
 D = {3 \over 2} R,
\end{eqnarray}
the anomalous dimension is obtained to be
\begin{eqnarray}
 \gamma_M = {1 \over N_f},\ \ \ 
\gamma_{q} + \gamma_{\bar q} = -{1 \over N_f}.
\end{eqnarray}
Therefore, with a large $N_f$ this dual description is a weakly coupled
theory.
Compared with the one-loop level computation:
\begin{eqnarray}
 \gamma_M = {h^2 N \over (4 \pi)^2}, \ \ \ 
 \gamma_{q} = \gamma_{\bar q} = {h^2 N_f \over (4 \pi)^2}
- {g^2 \over (4\pi)^2} {{N^2 - 1} \over N},
\end{eqnarray}
the critical values of $h$ and the gauge coupling $g$ are obtained to be
\begin{eqnarray}
 h_* = {4 \sqrt 3 \pi \over N_f} + O(1/N_f^2), 
\ \ \ 
 g_* = {4 \pi \over N_f} \sqrt{21\over 2} + O(1/N_f^2).
\label{eq:h}
\end{eqnarray}
With these values, the loop expansion parameters are
\begin{eqnarray}
 {h_*^2 N \over (4 \pi)^2} = {1 \over N_f} + O(1/N_f^2),\ \ \ 
 {g_*^2 N \over (4 \pi)^2} = { 7 \over 2 N_f} + O(1/N_f^2).
\end{eqnarray}
Therefore, a systematic loop expansion is possible for a large $N_f$.

\section{Soft SUSY breaking terms}

In the dual description, we expect that there are a gaugino mass for the
$SU(N)$ gauge group, the $A$-term for the superpotential, and the soft
scalar masses for the dual quarks and the meson
superfields~\cite{Aharony:1995zh}.
We discuss below the renormalization group (RG) evolution of the soft SUSY
breaking terms.
See~\cite{ArkaniHamed:1998wc, Karch:1998qa, Luty:1999qc} for exact
results and sum rules among soft terms in SUSY QCD theories.
In particular, in Refs.~\cite{Karch:1998qa, Luty:1999qc} the RG
evolutions of the soft terms near the IR fixed points have been studied
and it has been observed that all the soft terms eventually flow to zero
if certain conditions are satisfied.

The essential observation of Ref.~\cite{Luty:1999qc} is that the soft
SUSY breaking terms in the electric picture can be recast in the form of
the modified anomaly mediation~\cite{Randall:1998uk, Giudice:1998xp}. In
the case where $Q$ and $\bar Q$ have common scalar masses, there are
only two parameters: the scalar mass, $\tilde m_Q^2$, and the gaugino
mass, $m_\lambda$.
Those two parameters can be thought of as the ones obtained from the
$F$-term in the anomaly mediation and a $D$-term of $U(1)_R$
symmetry. It has been shown in~\cite{Pomarol:1999ie} that such a
$D$-term modification just results in a simple deformation of the RG
trajectory if the $R$-symmetry is not anomalous. The modified
trajectories~\cite{Luty:1999qc} are given by
\begin{eqnarray}
 \tilde m_Q^2 (\mu) = \left(
{2 \over 3}(1 + \gamma_Q ) - R_Q
\right) D_R
+ {1 \over 2} \dot \gamma_Q |F_\phi|^2,
\end{eqnarray}
\begin{eqnarray}
 m_\lambda ( \mu ) = {\beta (g^2) \over 2 g^2 } F_\phi,
\end{eqnarray}
where $F_\phi$ and $D_R$ are the $F$-term of the conformal compensator
and the $U(1)_R$ $D$-term, respectively, and $R_Q$ is the $U(1)_R$
charge of $Q$ and $\bar Q$. Once the soft terms are described in the
above form with some values of $F_\phi$ and $D_R$ at some scale, soft
parameters at any scales are obtained by the above formula. In
particular, at the conformal fixed point, the soft terms vanish, and
SUSY is recovered.

Even in a more general case where the soft scalar masses are flavor
dependent, the solution of the RGE is still very simple. The deviation
from the common scalar masses can always be parametrized as
\begin{eqnarray}
 \Delta \tilde m_{Q_i}^2 = q_i^a D^a,
\end{eqnarray}
where $q_i^a$ and $D^a$ are the charges and $D$-terms of $U(1)$ subgroups
of the anomaly-free flavor symmetry.
In this case, the RG trajectory is simply modified to
\begin{eqnarray}
 \tilde m_{Q_i}^2 (\mu) = \tilde m_Q^2 (\mu) + q_i^a D^a,
\end{eqnarray}
as shown in Ref.~\cite{ArkaniHamed:2000xj}. Therefore, at the conformal
fixed point, the scalar masses are 
\begin{eqnarray}
 \tilde m_{Q_i}^2 = q_i^a D^a.
\end{eqnarray}
In summary, in the electric picture, the gaugino mass vanishes and the
scalar masses are sum of the contributions which are proportional to
global $U(1)$ charges, {\it i.e.,} the external $D$-terms.

This suggests that, in the magnetic picture, the gaugino mass and
$A$-term vanish and the soft masses for the dual quarks and meson fields
are just sums of $D$-terms. In this case, there are sum rules among
scalar masses of the meson fields,
\begin{eqnarray}
 \tilde m_{X_{\bf 1}}^2 = \tilde m_{X_{\rm adj.}}^2,
\label{eq:sum-1}
\end{eqnarray}
\begin{eqnarray}
\tilde m_{H_d}^2 - \tilde m_{H_u}^2 = 
\tilde m_{t^c}^2 - \tilde m_{b^c}^2,
\label{eq:sum-2}
\end{eqnarray}
\begin{eqnarray}
 \tilde m_{H_d}^2 + \tilde m_{X_{\bf 1}}^2 = 
 \tilde m_{q_3}^2 + \tilde m_{t^c}^2,
\label{eq:sum-3}
\end{eqnarray}
\begin{eqnarray}
 m_{H_u}^2 + m_{H_d}^2 + 3 m_{X_{\bf 1}}^2 = 0.
\label{eq:sum-4}
\end{eqnarray}

One can indeed see a fixed point structure at leading order in the
perturbation.
At $O(1/N_f)$, the RG equations for the gaugino mass ($m_{1/2}$),
$A$-term ($A_h$) and the scalar masses ($\tilde m_M^2$, $\tilde m_q^2$
and $\tilde m_{\bar q}^2$) are
\begin{eqnarray}
 {d m_{1/2} \over d \ln \mu} = O \left( 1 / N_f^2 \right),
\end{eqnarray}
\begin{eqnarray}
 {d A_h \over d \ln \mu} = {14 \over N_f} (A_h - m_{1/2}),
\end{eqnarray}
\begin{eqnarray}
 {d \tilde m_{M_{ij}}^2 \over  d \ln \mu} = {2 \over N_f} \left(
\tilde m_{M_{ij}}^2 + \tilde m_{q_i}^2
  + \tilde m_{\bar {q}_j}^2 + A_h^2 \right),
\end{eqnarray}
\begin{eqnarray}
 {d \tilde m_{q_i}^2 \over  d \ln \mu}
= {6 \over N_f^2} \sum_j \left(
\tilde m_{M_{ij}}^2 + \tilde m_{q_i}^2
  + \tilde m_{\bar q_j}^2 + A_h^2 \right)
- {14 \over N_f} m_{1/2}^2.
\end{eqnarray}
\begin{eqnarray}
 {d \tilde m_{\bar q_j}^2 \over  d \ln \mu} 
= {6 \over N_f^2} \sum_i \left(
 \tilde m_{M_{ij}}^2 + \tilde m_{q_i}^2
  + \tilde m_{\bar q_j}^2 + A_h^2 \right)
- {14 \over N_f} m_{1/2}^2.
\end{eqnarray}
We can find attractive IR fixed points for combinations of the soft
terms:
\begin{eqnarray}
 \tilde m_{M_{ij}}^2 + \tilde m_{q_i}^2
  + \tilde m_{\bar q_j}^2 = m_{1/2}^2,
\label{eq:fp-scalar}
\end{eqnarray}
\begin{eqnarray}
 A_h = m_{1/2}.
\label{eq:fpsoft3}
\end{eqnarray}
 From Eq.~(\ref{eq:fp-scalar}), Eqs.~(\ref{eq:sum-1}), (\ref{eq:sum-2}), and
 (\ref{eq:sum-3}) can be derived, {\it i.e.,} the scalar masses can be
 described by $D$-terms including $D_R$.
Since the gaugino mass $m_{1/2}$ eventually approaches to zero by the
sub-leading running effect, the $A$-term also vanishes. The $D_R$-term
contribution should also approach to zero by repeating the discussion of
the modified anomaly mediation in the dual picture. This leads
Eq.~(\ref{eq:sum-4}).

It is interesting to note here that there is no logarithmic enhancement
in $m_{H_u}^2$ to cause the fine-tuning problem due to the
non-renormalized nature of $D$-terms.

\section{Dynamical Electroweak Symmetry breaking}

The dynamical electroweak symmetry breaking in the electric picture can
be described as the Higgs mechanism in the dual magnetic picture. We see
in this section that the meson fields obtain VEVs by the help of the
soft SUSY breaking terms.

\subsection{The Higgs potential}
At a vacuum where $M$ obtains a VEV, the dual quarks become massive. By
integrating out the dual quarks we obtain a non-perturbatively generated
superpotential~\cite{Affleck:1983mk},
\begin{eqnarray}
 W_{\rm eff} = \kappa N  \left(
{h^{N_f} \det M \over \Lambda_*^{N_f-3N}}
\right)^{1/N},
\label{eq:superp}
\end{eqnarray}
where
\begin{eqnarray}
 \kappa \equiv e^{-{8 \pi^2 \over g_*^2 N}} = e^{-N_f/7 + O(1)}.
\label{eq:kappa}
\end{eqnarray}
The scale $\Lambda_*$ in the superpotential is an arbitrary scale below
$\Lambda$. Here, we canonically normalize the kinetic term of $M$ at the
scale $\Lambda_*$ so that the holomorphic and the physical gauge
coupling coincide. The K{\" a}hler potential is, therefore,
\begin{eqnarray}
 K = \left( \mu \over \Lambda_* \right)^{-2\gamma_M} M^\dagger M,
\label{eq:kahler}
\end{eqnarray}
where $\gamma_M = 1/N_f$. The $\Lambda_*$ dependence disappears from the
Lagrangian when the meson fields are canonically normalized.

One can use the above effective potential to look for a vacuum as long
as
\begin{eqnarray}
 h \langle M \rangle \gg m_{\rm SUSY}.
\label{eq:cond}
\end{eqnarray}
The perturbative correction to the potential (the Coleman-Weinberg
potential) in this regime is obtained by using the running soft masses
at the scale $\mu = h \langle M \rangle$. (There is no running of scalar
masses once soft terms approaches to the fixed point.)

The gauge invariance allows us to write down a mass term for the lower
three components of $Q$ and $\bar Q$ in the electric description. By
including the mass term, the corresponding operator in the magnetic
description is
\begin{eqnarray}
 W_{\rm mass} = \mu_X^2 \tr X,
\end{eqnarray}
where $\mu_X$ is a new parameter of mass dimension one.

Putting $W_{\rm eff}$, $W_{\rm mass}$ and soft SUSY breaking terms
together, the potential for the scalar components of $M$ is given by
\begin{eqnarray}
 V &=& 
\sum_{i,j}
\left|
\kappa h^3 \left(
\det M
\right)^{3/N_f} 
M^{-1}_{ij}
+
\sum_I
(h \det M^{1/N_f})^{1/N_f}
\hat \mu_X^{2 - 1/N_f}
\delta_{iI} \delta_{jI}
\right|^2
\nonumber \\
&&
+(\tilde m_M^2)_{ij} \sum_{i,j} | M_{ij} |^2
+ \left(
(h \det M^{1/N_f})^{1/N_f}
\hat \mu_X^{2 - 1/N_f} A_X \tr X
+ {\mbox{h.c.}}
\right)
.
\label{eq:pote}
\end{eqnarray}
The wave function is normalized at $\mu = (h^{N_f} \det \langle M
\rangle)^{1/N_f}$. 
The index $I$ runs for flavors corresponding to $X$. The mass parameter
$\hat \mu_X$ is defined so that the combination of $\mu^{1/N_f} \hat
\mu_X^{2-1/N_f}$ gives the running mass parameter.
The last two terms are the soft SUSY breaking terms.

There can be, in principle, an $A$-term associated with the
superpotential term $W_{\rm eff}$.
This term, however, approaches to zero in the renormalization group
running in the limit of the vanishing Standard Model gauge couplings.
The deviation due to the QCD coupling is suppressed by a two-loop factor
($\alpha_s^2 / (4 \pi)^2$) compared to the gaugino masses. We therefore
ignore this $A$-term in the analysis.

Now let us examine whether there exists a stable vacuum where
electroweak symmetry is broken while satisfying Eq.~(\ref{eq:cond}). As
we will see later, the value of $\tan \beta \equiv \langle H_u \rangle /
\langle H_d \rangle$ is required to be close to unity considering the
constraint from the $T$-parameter. Therefore we here make an ansatz:
\begin{eqnarray}
\langle M \rangle = \left(
\begin{array}{c|c}
 v_H {\bf 1} & \\ \hline
 & v_X {\bf 1} \\
\end{array}
\right).
\end{eqnarray}
The $H_u$ and $H_d$ fields are treated as a single field $H = (H_d,
H_u)$ in the following discussion.
In order for the stop/sbottom fields not to be tachyonic, we require
\begin{eqnarray}
 \tilde m_{q_3}^2 >0, \quad \tilde m_{t^c}^2 > 0.
\end{eqnarray}
The requirement, $\tan \beta \simeq 1$, suggests that
\begin{eqnarray}
 \tilde m_{H_u}^2 = \tilde m_{H_d}^2 \equiv m_H^2.
\end{eqnarray}
Then the sum rules in Eqs.~(\ref{eq:sum-3}) and (\ref{eq:sum-4}) result in
\begin{eqnarray}
 \tilde m_{H}^2 > 0,\ \ \ \tilde m_{X_{\bf 1}}^2 < 0.
\end{eqnarray}

If we ignore the SUSY breaking terms, we find that the potential has a
runaway behavior in the $X \to 0$ and $H \to \infty$ direction.
Therefore, once we turn on the SUSY breaking terms with $\tilde m_H^2 >
0$ and $\tilde m_{X_{\bf 1}}^2 < 0$, there should be a vacuum at which
both $X$ and $H$ have finite values.

Taking $v_H \sim v_X \sim v$ and $\tilde m_H \sim \tilde m_X \sim A_X
\sim m_{\rm SUSY}$, the minimization of the potential leads
\begin{eqnarray}
 m_{\rm SUSY} \sim \kappa h^3 v,
\label{eq:eps}
\end{eqnarray}
and
\begin{eqnarray}
 \mu_X^2 \sim \kappa h^3 v^2.
\end{eqnarray}
The assumption $v_H \sim v_X$ can be relaxed, but a large deviation
causes a strong coupling at the vacuum with which the description in
terms of the Higgs fields does not make sense. One should integrate out
dual quarks in two steps in this case.

By comparing Eqs.~(\ref{eq:cond}) and (\ref{eq:eps}), one finds that the
analysis is reliable when
\begin{eqnarray}
 \kappa h^2 \ll 1.
\end{eqnarray}
In terms of $N_f$ this condition corresponds to 
\begin{eqnarray}
 N_f \gg 10.
\end{eqnarray}
Unfortunately, the minimal model ($N_f = 5$) is not quite in the
calculable regime.
Nevertheless, we proceed the discussion in order to see the qualitative
features of the model. We will discuss more on a case with $N_f = 5$
later.

The $F$-component of $H$ at the vacuum is
\begin{eqnarray}
 F_{H}^\dagger = - {\partial W_{\rm eff} \over \partial H} 
 \sim \kappa h^3 v^2. 
\end{eqnarray}
In the original electric picture, this quantity corresponds to the top
(bottom) condensations,
\begin{eqnarray}
 F_{H} \sim 
 \langle \psi_Q \psi_{\bar Q} \rangle,
\label{eq:F}
\end{eqnarray}
where $\psi_Q$ and $\psi_{\bar Q}$ are the fermionic components of $Q$
and $\bar Q$, respectively. The strong $SU(N_c)$ interaction gives a
large negative anomalous dimension for the fermion pair. The dimension
is $D(\psi_Q \psi_{\bar Q}) = D(F_M) = 2 + 1/N_f$.

Note however that this fermion-pair condensations are minor contributions
to the electroweak symmetry breaking for a large $N_f$ due to the
suppression factor, $\kappa h^3$. 
Moreover, $F_M$ is a composite operator also in the magnetic
description. Since there is no kinetic term for $F_M$, the VEV does not
directly contribute to the $W$ boson mass.
The main contribution to the $W$ boson mass is $\langle M \rangle$, {\it
i.e.,} the VEV of the Higgs field which corresponds to a boson-pair
condensation in the electric picture.

We stress here that the Higgs fields are stabilized by the
non-perturbative superpotential, not by the quartic terms from the
$D$-terms. Therefore, the Higgs boson mass is unrelated to the $Z$ boson
mass. Together with the absence of the logarithmic enhancement of the
soft terms, there is no tension between the experimental lower bound on
the Higgs boson mass and the fine-tuning in the electroweak symmetry
breaking.

\section{Fermion mass generation}

 Through the dynamical superpotential, the fermionic components of the
 meson fields, including the top quark, obtain masses, $|m_F|$, with
\begin{eqnarray}
 m_F^{ij,kl} = {\partial^2 W_{\rm eff} \over \partial M_{ij}
  \partial M_{kl}}
= \kappa h^3 (\det M)^{3/N_f}
\left(
{1 \over N} (M^{-1})_{ji} (M^{-1})_{lk}
- (M^{-1})_{li} (M^{-1})_{jk}
\right).
\label{eq:fermion}
\end{eqnarray}
For $N_f = 5$, the top and bottom quark masses are
\begin{eqnarray}
 m_t = - {\Lambda_L^3 \over \langle H_d \rangle \langle X \rangle},\ \ \  
 m_b = - {\Lambda_L^3 \over \langle H_u \rangle \langle X \rangle},
\label{eq:tbmass}
\end{eqnarray}
where $\Lambda_L^3 \equiv \kappa h^3 (\det \langle M \rangle)^{3/N_f}$.
The charged Higgsino and the color adjoints have masses:
\begin{eqnarray}
 m_{\tilde h^+} = 
- {\Lambda_L^3 \over \langle H_u \rangle \langle H_d \rangle},\ \ \  
 m_{X_{\rm adj.}} = 
- {\Lambda_L^3 \over \langle X \rangle^2}.\ \ \  
\end{eqnarray}
The neutral Higgsinos and the singlino have a mass matrix:
\begin{eqnarray}
 m_{\chi^0} = \left(
\begin{array}{ccc}
 -{\Lambda_L^3 \over 2 \langle H_d \rangle^2} 
& {\Lambda_L^3 \over 2 \langle H_u \rangle \langle H_d \rangle}
& {\sqrt{3} \over 2}{\Lambda_L^3 \over 
\langle H_d \rangle \langle X \rangle}\\
& -{\Lambda_L^3 \over 2 \langle H_u \rangle^2} 
& {\sqrt{3} \over 2}{\Lambda_L^3 
\over \langle H_u \rangle \langle X \rangle}\\
& 
& {\Lambda_L^3 \over 2 \langle X \rangle^2} \\
\end{array}
\right).
\end{eqnarray}

The top quark mass is proportional to the fermion-pair condensation:
$m_t \sim F_{H_d}^\dagger / \langle X \rangle$. With Eq.~(\ref{eq:F}),
we see that the picture of the dynamical top mass through the top
condensation emerges.

\section{Towards a realistic model}

We have seen in the previous section that the overall picture is quite
successful for a large $N_f$. The chiral symmetry breaking occurs and
the dynamical masses for the fermions are generated.

In order to apply this framework for the actual electroweak symmetry
breaking and the top-quark mass generation, we need to look at the model
more closely. Below, we list the questions to be considered.

\subsection{Calculability}

We have seen that $N_f = 5$ is not quite large enough to be in the
regime of $h \langle M \rangle \gg m_{\rm SUSY}$ where the calculation
is reliable.
Nevertheless, even with $N_f = 5$ it is justified to use a $1/N_f$
expansion and to identify the meson fields as weakly coupled effective
degrees of freedom. The problem is the use of the non-perturbative
superpotential in Eq.~(\ref{eq:superp}) near $M = 0$ where the soft
SUSY breaking becomes important.

Here we simply assume that the qualitative picture of a large $N_f$
theory persists for $N_f = 5$.
Namely, we assume that the electroweak symmetry breaking is described by
condensations of the meson fields, where the meson potential is given by
Eq.~(\ref{eq:pote}) with the coefficients $\kappa h^3$ treated as a free
parameter of order unity.
In this case, we have
\begin{eqnarray}
m_{\rm SUSY} \sim \langle H_{u,d} \rangle \sim m_t,
\end{eqnarray}
which are desired relations.

\subsection{Masses for light flavors}

At this point, all the leptons and the quarks in the first and second
generations are massless.
Those light fermions can obtain masses through higher dimensional
operators. For example, the charm quark mass arises from
\begin{eqnarray}
 W \ni {1 \over \Lambda_{\rm ETC}} q_2 u^c_2 (Q \bar Q)_{H_u} ,
\end{eqnarray}
where $(Q \bar Q)_{H_u}$ is the $SU(N_c)$ singlet combination with the
quantum number of $H_u$. 
Unlike the technicolor theories, the fermion masses obtained from those
terms are not simply suppressed by the factor of ${\rm TeV} /
\Lambda_{\rm ETC}$, because of the large anomalous dimension of the
operator $(Q \bar Q)_{H_u}$. With the dimension of the scalar component
of the operator, $1 + 1/N_f$, the charm mass $m_c$ is
\begin{eqnarray}
 m_c = {\Lambda \over \Lambda_{\rm ETC}} 
\left(
{h \bar v \over \Lambda}
\right)^{1/N_f} v_u,
\end{eqnarray}
where $\bar v \equiv (\det \langle M \rangle )^{1/N_f}$ and $\Lambda$
the scale where the theory flows into the fixed point.
With $N_f = 5$, one finds
\begin{eqnarray}
 \Lambda_{\rm ETC} = 3 \times 10^{12}~{\rm GeV}
\left(
h \bar v \over 300~{\rm GeV}
\right)
\left(
\Lambda \over \Lambda_{\rm ETC}
\right)^4.
\end{eqnarray}
The ETC scale can be postponed up to about $10^{12}$~GeV. The FCNC
problem is absent with such a high scale.

\subsection{Bottom quark mass}

If the fermion masses are dominated by the contribution from the
dynamical superpotential, we have from Eq.~(\ref{eq:tbmass}):
\begin{eqnarray}
 {m_t \over m_b} = {\langle H_u \rangle \over \langle H_d \rangle}  
\equiv \tan \beta.
\end{eqnarray}
On the other hand, there is a relation
\begin{eqnarray}
 m_t m_b = m_{{\tilde h}^+} m_{X_{\rm adj.}}
\end{eqnarray}
Therefore, a too large value of $\tan \beta$ results in a prediction of
either a very light Higgsino or a color adjoint fermion.
Moreover, as we will see later, a large deviation from $\tan \beta = 1$
is dangerous for the $T$-parameter constraint.

A possibility to cure this unacceptable situation is to extend the model
as in the original topcolor model in Ref.~\cite{Hill:1991at}. There, an
$SU(N_c)$ singlet vector-like pair $b^\prime: (3,1)_{-1/3}$ and
$b^{\prime c}: (\bar 3, 1)_{1/3}$ is introduced. By writing a
superpotential term, $(Q \bar Q)_{b^c} b^\prime$, the bottom quark
remains massless after electroweak symmetry breaking. A term $b^\prime
b^{\prime c}$ with a small coefficient induces a small mass for the
$b$-quark.

\section{$S$ and $T$ parameters}

The dual quarks obtain masses from the electroweak symmetry
breaking. The loop diagrams with the dual quarks, therefore, contribute
to the $S$ and $T$ parameters~\cite{Peskin:1990zt}.
For small SUSY breaking parameters, the main contribution to $S$ is from
a fermion loop:
\begin{eqnarray}
 \Delta S = {N \over 6 \pi}.
\end{eqnarray}
With $N=2$, it is $\Delta S \simeq 0.1$. This size of contribution is
allowed if there is a light Higgs boson~\cite{Erler:2010sk}.

If we ignore the soft terms, the contribution to the $T$ parameter is
\begin{eqnarray}
 \Delta T = {N \over 8 \pi s^2 c^2 m_Z^2}
\left[
m_1^2 + m_2^2 - {2 m_1^2 m_2^2 \over m_1^2 - m_2^2} \log {m_1^2 \over m_2^2}
\right],
\end{eqnarray}
where $m_1$ and $m_2$ are supersymmetric masses for the $SU(2)_L$
doublet dual quarks,
\begin{eqnarray}
 m_1 = h \langle H_d \rangle, \ \ \ m_2 = h \langle H_u \rangle.
\end{eqnarray}
For a large $N_f$, one can rewrite this to
\begin{eqnarray}
 \Delta T &=& {1 \over \alpha N_f}
\left[
1 + {\sin^2 2 \beta \over \cos 2 \beta} \log \tan \beta
\right]
\\
&=&
{2 \over 3 \alpha N_f}
\left[
(\tan \beta -1 )^2 + O((\tan \beta - 1)^3)
\right]
\end{eqnarray}
at leading order in $1/N_f$. The contribution vanishes when $\tan \beta
= 1$. For other values of $\tan \beta$,
\begin{eqnarray}
 \Delta T = 0.15 \left(5 \over N_f\right)\ \ \ (\tan \beta = 1.1),
\end{eqnarray} 
\begin{eqnarray}
 \Delta T = 0.57 \left(5 \over N_f\right)\ \ \ (\tan \beta = 1.2).
\end{eqnarray} 
Although the estimation is not very reliable for $N_f = 5$, a large
deviation from $\tan \beta = 1$ would not be preferred.

\section{Discussion and conclusions}

We have considered a SUSY QCD model ($N_c = 3$ and $N_f = 5$) with soft
SUSY breaking terms as a model for dynamical electroweak symmetry
breaking.  The theory is in a conformal window and has a dual
description in terms of the dual quarks and the meson superfields.  The
top/bottom quarks and the Higgs fields are embedded in the meson
superfield.

We find at a large $N_f$, a perturbative expansion is reliable and we
can see that there exists a stable vacuum with non-zero VEVs for the
Higgs operator (the stop condensation).
The top/bottom quarks and the Higgsinos obtain masses through the
dynamically generated superpotential. 
Thanks to the conformal regime of the dynamics, light flavors can obtain
masses without introducing the FCNC problem.
The low energy effective theory looks like the MSSM but the $\mu$-term
and the top/bottom Yukawa interactions replaced by a dynamical
superpotential, where fractional powers of the superfields appear as a
consequence of integrating out the dual quarks.  
The effective superpotential is
\begin{eqnarray}
 W = W_{\rm Yukawa} + 
\Lambda_*^{1/2} \left(
X^3 H_u H_d + q_3 t^c H_u X^2 + q_3 b^c H_d X^2 +
  q_3 t^c q_3 b^c X 
\right)^{1/2} + \mu_X^2 X,
\end{eqnarray}
and the K{\" a}hler potential is given in Eq.~(\ref{eq:kahler}).  The
first term, $W_{\rm Yukawa}$, is the Yukawa interaction terms for
leptons and first and the second generation quarks.

There are numbers of interesting phenomenology which we reserve for
future studies.
There are a color adjoint fermion and a boson ($X_{\rm adj.}$) with
masses at the electroweak scale.
Unlike the gluino, the color adjoint fields couple only to the
top/bottom (s)quarks. Therefore, the boson and the fermion parts decay
into $b \bar b$ and $b \bar b \chi^0$, respectively.
An LHC study of such particles can be found in~\cite{Gerbush:2007fe}.
There are also colored components of the dual quarks which are charged
under the strong $SU(2)$ dual gauge group. Therefore, hadrons of the
strong $SU(2)$ dynamics may also be observable at the LHC.

The Higgs and the top/bottom sectors are quite different from those of
the MSSM.
There are two neutral Higgs fields (the real and imaginary parts of $X$)
in addition to $h$, $H$, and $A$ in the MSSM. Since the parameters
$\mu_X$ and $A_X$ explicitly break $U(1)_{PQ}$ and $U(1)_{R}$,
respectively, both new Higgs fields obtain masses of the order of the
electroweak scale.
The term with the fractional power in the superpotential implies that
the fermion masses and the Yukawa coupling constants are in general not
aligned, as is always the case in the topcolor model.
Especially, the top Yukawa coupling (the coupling to the lightest Higgs
boson) is not the same as $m_t / \langle H \rangle$, where $\langle H
\rangle$ is the Higgs VEV measured by the Fermi constant.
This will be interesting for $B$ physics and also for the top physics.

\section*{Note Added}
After this paper was written, J.~Evans {\it et~al.} put out a paper
which discusses the supersymmetric topcolor model~\cite{Evans:2010ed} as
an ultraviolet completion of the conformal technicolor.

\section*{Acknowledgements}
RK thanks the organizers of the KITP workshop ``Strings at the LHC and
in the Early Universe,'' March 8 -- May 14, 2010, where this work was
initiated, and Yutaka Ookouchi for useful discussions at the workshop.
RK is supported in part by the Grant-in-Aid for Scientific Research
21840006 of JSPS.

\end{document}